\documentclass[prd,onecolumn,12pt,tightenlines,nofootinbib]{revtex4}
\usepackage{graphicx}

\newcommand{\im}{{\mathrm Im\,}}
\newcommand{\cl}{\underline{\lambda}}
\newcommand{\mh}{m_h}

\begin{document}

\title{Heavy-quarkonium hadron cross section in QCD at leading twist}

\author{Fran\c cois Arleo}
\author{Pol-Bernard Gossiaux}
\author{Thierry Gousset}
\author{J\"org Aichelin}
\affiliation{
SUBATECH, Laboratoire de Physique Subatomique et des Technologies
Associ\'ees\\
UMR Universit\'e de Nantes, IN2P3/CNRS, Ecole des Mines de Nantes\\
4, rue Alfred Kastler, 44070 Nantes cedex 03, France}

\begin{abstract}
We compute the total cross section of a heavy quarkonium on a hadron
target in leading twist QCD, including target mass corrections. Our
method relies on the analytical continuation of the operator product
expansion of the scattering amplitude, obtained long ago by Bhanot and
Peskin. The cross section has a simple partonic form, which allows us
to investigate the phenomenology of $J/\psi$ and $\Upsilon$
dissociation by both pions and protons.
\end{abstract}


\maketitle

\section{Introduction}

It has been conjectured that in ultra-relativistic collisions between
heavy ions a quark gluon plasma (QGP) is formed for a very short
period of time. Later it disintegrates into the hadrons which are
finally seen in the detectors. It is the challenge of the present
experiments at the CERN-SPS and at the Relativistic Heavy Ion Collider
(RHIC) to find observables which unambiguously signal the formation of
the QGP.

One of the most promising suggestions advanced so far is the
suppression of charmonium production in these collisions. It has been
argued by Matsui and Satz~\cite{MS} that the interaction between a
heavy quark $Q$ and antiquark $\bar Q$ is screened in a QGP, and
consequently the $Q\bar Q$ bound pairs may not survive in this
environment. The problem with this signal is that all other possible
suppression mechanisms have to be well understood.

Whereas the general features of charmonium production in proton-proton
collisions seem to be under control, already in proton-nucleus
reactions the suppression is not well understood so far. Only
recently, data have been published which show a different suppression
of $J/\psi$ and $\psi'$~\cite{Le}, and hence give the first hints
that the $J/\psi$ is formed inside the nucleus.

In heavy ion collisions the situation is even more difficult. There
many particles are produced which possibly collide with a charmonium
and may cause an observable suppression even if a QGP is not formed at
all. In order to quantify such a suppression it is necessary to know
the strength of these interactions. Experimentally the
charmonium-hadron dissociation cross sections are not
accessible. Therefore one has to rely on theoretical estimates. Three
kinds of approaches have been advanced in the past.

The first approach is based on twist expansion techniques well known
from deep inelastic scattering studies. It has been launched by Bhanot
and Peskin~\cite{Pe,BP} and was explicitly used in Ref.~\cite{BP} for
$\psi$-$p$ in the approximation of a vanishing proton mass. The whole
approach gives a correct approximation of QCD provided that the heavy
quark mass is large enough. It has thus the very advantage to be a
well defined approximation scheme of the underlying theory. For
realistic systems, such as charmonia and bottomonia, power corrections
are presumably non negligible, but some of them, namely finite target
mass corrections, can be incorporated in a systematic way.

A second attempt is formulated within a constituent quark model. In an
early study Martins \textit{et al.}~\cite{Ma} have calculated a
$J/\psi$ dissociation cross section $\sigma_{\psi\,\pi}$ by $\pi$'s of
up to 7~mb at $\sqrt{s}=4$~GeV, i.e., 0.8~GeV above
$m_{J/\psi}+m_\pi$. This value has been reduced to about 1~mb for the
same energy in a more recent study by Wong \textit{et al.}~\cite{Wo}
who used parameters adjusted to other elementary reactions.

A third approach is based on hadronic degrees of freedom. Invoking a
local U(4) symmetry and employing pseudoscalar-pseudoscalar-vector
coupling, Matinyan and M\"uller~\cite{MM} investigated the
dissociation of the $J/\psi$ by exchange of a $D$ or $\bar D$
meson. Employing vector dominance to determine the coupling constants
they arrive at $\sigma_{\psi\,\pi}\approx 0.3$~mb for
$\sqrt{s}=4$~GeV. Later, Haglin~\cite{Ha} included four point
interactions and a three vector-meson coupling and obtained a much
larger cross section because the large suppression of the cross
section due to the $D$-meson propagator is not present in the contact
terms. Recently Lin and Ko~\cite{Ko} modified the details of this
approach and included form factors. Depending on the form factor
assumed they get $\sigma_{\psi\,\pi}$ between 4 and 25~mb at
$\sqrt{s}=4$~GeV.

The purpose of this article is to extend the work Ref.~\cite{BP} in
three different directions. First of all we include systematically the
masses of the scattering partners using a method known from deep
inelastic scattering studies. This allows for the calculation of the
dissociation cross sections close to threshold where it is most
relevant for the question at hand. Second the cross section of
Ref.~\cite{BP} is derived in a different, more direct fashion, again
in analogy to the calculation of the forward Compton scattering
amplitude in the operator product expansion. This cross section has a
simple partonic expression, even when target mass corrections are
included. We also explicitly derive how the reaction threshold is
shifted by target mass corrections and how the cross section is
modified in the vicinity of threshold. Third the calculation is
extended towards other hadrons $h$ and towards bottomonium which becomes
an observable particle in the upcoming experiments at the RHIC and at
the Large Hadron Collider at CERN.

Target mass correction for $J/\psi$-$p$ cross section has been
examined in the framework of Bhanot and Peskin in
Ref.~\cite{KSSZ}. These authors obtained their results in the form of
sum rules. In principle these sum rules contain all of the above
mentioned aspects but none is made explicit. Further we were neither
capable to reproduce their exact expressions for the sum rules nor to
find the trend they mention for the correction. We will clarify in the
course of this study where we disagree.

\section{Derivation}\label{sec:derivation}

In this section we generalize the expression obtained by Bhanot and
Peskin~\cite{BP} for the total cross section of a heavy quarkonium
$\Phi$ with a target hadron $h$ by including finite target mass
terms. The proposed analysis is close to that performed in the context
of deep inelastic scattering~\cite{GP,Na}.

\subsection{Short review of the framework}\label{sub:framework}

Let us first collect the material we need from Refs.~\cite{Pe,BP}. We
want to compute the $\Phi$-$h$ total cross section\footnote{Throughout
this paper we restrict ourselves to \emph{spin-averaged} cross
sections without further mention.}. Our starting point is the
expression for the forward $\Phi$-$h$ elastic scattering amplitude,
$\mathcal{M}_{\Phi\,h}$. This amplitude depends on energy and it is
convenient to express it in terms of
\[
\lambda=\frac{(K+p)^2-M^2-\mh^2}{2M},
\]
where $K$ and $p$ are the $\Phi$ and $h$ respective 4-momenta and $M$
and $m_h$ their respective masses. We note that $\lambda$ is the
hadron energy in the $\Phi$ rest frame. Via the optical theorem, the
forward scattering amplitude leads to the $\Phi$-$h$ total cross
section
\[
\sigma_{\Phi\,h}(\lambda)=\frac{1}{\sqrt{\lambda^2-\mh^2}}
\im\mathcal{M}_{\Phi\,h}(\lambda).
\]
Notice that we use the same definition for $\mathcal{M}$ as in
Ref.~\cite{BP}. 

In QCD, in the limit of a large heavy quark mass, the scattering
amplitude has a twist expansion. In the $\Phi$ rest frame the Leading
Twist (LT) contribution is~\cite{BP}
\begin{equation}\label{eq:lt_amplitude}
\mathcal{M}_{\Phi\,h}^{(\mathrm{LT})}(\lambda)=a_0^3\epsilon_0^2
\sum_{k\ge 1}d_{2k}\epsilon_0^{-2k}
\langle h|F^{0\nu}(iD^0)^{2k-2}F^{\phantom{\nu}0}_\nu|h\rangle.
\end{equation}
$a_0$ and $\epsilon_0$ are, respectively, the Bohr radius and the
Rydberg energy for the $Q\bar Q$ system.

The above formula displays the factorization of the process in terms
of hard coefficients $d_n$ and soft matrix elements. Both should be
evaluated at a factorization (and renormalization) scale $\mu$, to be
chosen to minimize the influence of neglected higher order
perturbative corrections. It is argued in Ref.~\cite{Pe} that
$\mu\sim\epsilon_0$, though a precise determination fulfilling the
latter requirement would need a complete one-loop computation. In the
phenomenological study, we will quantify the consequences of this
scale uncertainty.

The coefficients $d_n$ correspond to matrix elements of definite
operators evaluated in the $\Phi$-state. These are computable in
perturbative QCD and have been made explicit in Ref.~\cite{Pe} for 1S
and 2S $\Phi$-states to leading order in the coupling and to leading
order in $1/N_c$, where $N_c$ is the number of colors. For
1S-state\footnote{The present consideration applies to 2S-state as
well but is left out until Appendix~\ref{app:2S_states}} they read
\[
d_n=\frac{16^3}{3N_c^2}\,B(n+5/2,5/2),
\]
where $B(\mu,\nu)$ is the Euler beta function. For later convenience
we remark that $d_n$ can be expressed as the $n$-th moment of a given
function $f$ through
\[
d_n=\int_0^1 \frac{dx}{x} x^n f(x),
\]
with
\[
f(x)=\frac{16^3}{3N_c^2} x^{5/2}(1-x)^{3/2}.
\]

For every $k$ in Eq.~(\ref{eq:lt_amplitude}) a gluon twist-2 operator
evaluated in the hadron $h$ (\emph{spin-averaged}) state also
appears. Each of these matrix elements is a traceless fully symmetric
rank $2k$ tensor built from the hadron momentum $p^\mu$. It turns out
that a tensor having these properties is necessarily proportional
to~\cite{GP}
\begin{eqnarray*}
\Pi^{\mu_1\cdots\mu_{2k}}(p)&=&\sum_{j=0}^k (-m_h^2)^j
\frac{(2k-j)!}{2^j (2k)!}\\
&&\times\Biggl[\sum_{\mathrm{perm.}}
\underbrace{g\otimes\cdots\otimes g}_{j}\otimes
\underbrace{p\otimes\cdots\otimes p}_{2k-2j}\Biggr]^{\mu_1\cdots\mu_{2k}},
\end{eqnarray*}
where the tensor in the $j$th term on the right-hand side is the sum
of the $(2k)!/[2^j j! (2k-2j)!]$ distinct tensors one can construct
by multiplying $j$ $g^{\nu\rho}$'s and $(2k-2j)$ $p^{\sigma}$'s. The
matrix element needed in Eq.~(\ref{eq:lt_amplitude}) is therefore
proportional to $\Pi^{0\cdots 0}(p)$ and writes
\begin{eqnarray}
&&\langle h(p)|F^{0\nu}(iD^0)^{2k-2}F^{\phantom{\nu}0}_\nu|h(p)\rangle
=A_{2k}\,\Pi^{0\cdots 0}(p)\nonumber\\
\label{eq:gluon_matrix_element}
&&=A_{2k}\sum_{j=0}^k \frac{(2k-j)!}{j!(2k-2j)!} 
(-m_h^2/4)^j \lambda^{2k-2j}. 
\end{eqnarray}
This set of matrix elements is related to the unpolarized gluon
density $G$ in the hadron target. One can see this from the matrix
element definition of $G$, see, e.g.,~\cite{CTEQ}, which in the
light cone gauge $A^+=0$ reads
\[
xG(x)=\frac{1}{p^+}\int\frac{dy^-}{2\pi}e^{ixp^+y^-}
\langle h(p)|F^{+\nu}(0)F^{\phantom{\nu}+}_\nu(y^-)|h(p)\rangle.
\]
In the parton model the argument of $G$, i.e., $x$, is interpreted as
the fraction of the hadron light cone momentum $p^+$ carried by the
gluon. Taking the $n$th moment of $G(x)$, we get
\[
\int_0^1 \frac{dx}{x} x^n G(x)=\frac{1}{(p^+)^n}\langle h(p)|
F^{+\nu}(i\partial^+)^{n-2}F^{\phantom{\nu}+}_\nu|h(p)\rangle.
\]
Since $D^+=\partial^+$ in the $A^+=0$ gauge we recognize
\begin{eqnarray*}
\langle h(p)|F^{+\nu}(iD^+)^{2k-2}F^{\phantom{\nu}+}_\nu|h(p)\rangle
&=&A_{2k}\,\Pi^{+\cdots +}(p)\\
&=&A_{2k}\,(p^+)^{2k},
\end{eqnarray*}
i.e.,
\[
A_n=\int_0^1 \frac{dx}{x} x^n G(x).
\]

\subsection{Massless target}\label{sub:mh0}

In the present subsection we want to illustrate the general method
in the case of vanishing target mass, $\mh=0$. Then
Eq.~(\ref{eq:gluon_matrix_element}) simplifies to
\[
\langle h|F^{0\nu}(iD^0)^{2k-2}F^{\phantom{\nu}0}_\nu|h\rangle
=A_{2k}\lambda^{2k},
\]
and Eq.~(\ref{eq:lt_amplitude}) thus becomes
\begin{equation}\label{eq:power_series}
\mathcal{M}_{\Phi\,h}^{(\mathrm{LT})}(\lambda)=a_0^3\epsilon_0^2
\sum_{k\ge 1}d_{2k}A_{2k}(\lambda/\epsilon_0)^{2k}.
\end{equation}

It is useful to study the scattering amplitude $\mathcal{M}_{\Phi\,h}$
throughout the complex plane of the energy. To avoid confusion, we
will from now on reserve the notation $\lambda$ to real values and
define $\cl$ as the extension of $\lambda$ to complex values. Using
the d'Alembert criterion one easily checks that the convergence radius
of the power series~(\ref{eq:power_series}), now considered with the
complex argument $\cl$, is equal to $\epsilon_0$. As extensively
discussed in Ref.~\cite{BP}, the twist expansion of the scattering
amplitude provides an expression for
$\mathcal{M}_{\Phi\,h}^{(\mathrm{LT})}(\lambda)$ in the
\emph{unphysical} region of energies. Since we are interested in
\emph{physical} energies we have to perform an analytic continuation
of the power series.

Before doing so let us first elaborate on possible differences between
$\mathcal{M}_{\Phi\,h}^{(\mathrm{LT})}$ and the \emph{full} scattering
amplitude $\mathcal{M}_{\Phi\,h}$, i.e., including higher twist
terms. It turns out from the analysis below that the singularities
(branch points) of $\mathcal{M}_{\Phi\,h}^{(\mathrm{LT})}$ on the
boundary of the convergence disk of the power series lie at
$\cl=\pm\epsilon_0$. This is not what is expected for the locations of
the (first) branch points of $\mathcal{M}_{\Phi\,h}$, i.e., the
locations of the thresholds for both reactions $\Phi+h\to X$ and
$\Phi+\bar h\to X$, which are $\cl=\pm \mh$. The technical reason for
the difference is, of course, that $\mh$ occurs nowhere in
Eq.~(\ref{eq:power_series}). In the twist expansion approach the
locations of the singularities of the full scattering amplitude may be
affected by higher twist corrections. We verify that this is the case
for the above reactions by taking into account elastic unitarity
corrections for which the thresholds are clearly located at $\cl=\pm
\mh$.\footnote{Let us note in passing one important phenomenological
consequence of the LT analysis : what we call the \emph{total} (LT)
cross section does not in fact include processes such as the elastic
one. The word ``total'' is thus misleading, at least in the threshold
region.} In the next section we shall see how $\mh\neq 0$ corrections
affect the locations of the LT reaction thresholds.

Since the convergence radius of the power series
(\ref{eq:power_series}) is non zero the LT amplitude\footnote{From
now on $\mathcal{M}$ always refers to the leading twist part of the
forward $\Phi$-$h$ elastic scattering amplitude and we drop the
indices (LT) and $\Phi\,h$ for simplicity.} can be unambiguously
determined throughout the $\cl$ complex plane from the sole knowledge
of Eq.~(\ref{eq:power_series}), using the Mellin transform
machinery. We first remark that $M_n=d_n A_n$ being a product of
moments one can express
\begin{equation}\label{eq:moments}
M_n=\int_0^1 \frac{dx}{x} x^n h(x),
\end{equation}
with
\[
h(x)=G\otimes f(x),
\]
and the convolution product defined as
\[
G\otimes f(x)=\int_x^1 \frac{dy}{y} G(y)\,f(x/y).
\]
Now, plugging Eq.~(\ref{eq:moments}) in Eq.~(\ref{eq:power_series})
for $|\cl|<\epsilon_0$, freely interchanging summation with
integration and summing the ensuing geometrical series, one finds
\[
\mathcal{M}(\cl)=a_0^3\epsilon_0^2\int_0^1 \frac{dx}{x} h(x)
\frac{x^2(\cl/\epsilon_0)^2}{1-x^2(\cl/\epsilon_0)^2}.
\]
The key point is that this integral representation can be extended
throughout the entire complex plane except for the two branch points
$\cl=\pm\epsilon_0$. The analytic continuation of
Eq.~(\ref{eq:power_series}) to energies $\lambda>\epsilon_0$ is then
easily derived
\begin{equation}\label{eq:amplitude}
\mathcal{M}(\cl=\lambda\pm i\varepsilon)=a_0^3\epsilon_0^2
\int_0^1 dx\,h(x)\frac{x(\lambda/\epsilon_0)^2}
{1-x^2(\lambda/\epsilon_0)^2\mp i\varepsilon},
\end{equation}
and its imaginary part given by
\[
\im\mathcal{M}(\lambda)=\frac{1}{2i}[\mathcal{M}(\lambda+i\varepsilon)-
\mathcal{M}(\lambda-i\varepsilon)]
=\frac{\pi}{2}a_0^3\epsilon_0^2\,h(\epsilon_0/\lambda).
\]
Putting all things together, one gets
\begin{equation}\label{eq:im_amplitude}
\im{\cal M}(\lambda)=\frac{\pi\lambda}{2}a_0^3\epsilon_0
\frac{16^3}{3N_c^2}\int_{\epsilon_0/\lambda}^1 dx\,G(x)
\frac{(x\lambda/\epsilon_0-1)^{3/2}}{(x\lambda/\epsilon_0)^5}.
\end{equation}

Dividing Eq.~(\ref{eq:im_amplitude}) by $\lambda$ (the flux factor
when $\mh=0$) we recover the partonic expression of the $\Phi$-$h$
total cross section as obtained by Bhanot and Peskin within a parton
model approach, i.e.,
\begin{equation}\label{eq:cross_section}
\sigma_{\Phi\,h}(\lambda)=\int_0^1 dx\,G(x)\,\sigma_{\Phi\,g}(x\lambda),
\end{equation}
with the $\Phi$-gluon cross section
\begin{equation}
\sigma_{\Phi\,g}(\omega)=\frac{16^3\pi}{6N_c^2}a_0^3\epsilon_0
\frac{(\omega/\epsilon_0-1)^{3/2}}{(\omega/\epsilon_0)^5}
\theta(\omega-\epsilon_0),
\end{equation}
$\omega$ corresponding to the gluon energy in the $\Phi$ rest
frame. Aside from its energy dependence, the $\Phi$-gluon cross
section is driven by $a_0^3\epsilon_0\propto \alpha_S a_0^2$, as
expected in QCD for the interaction of a small color singlet dipole of
size $a_0$.

In this formulation one important physical aspect is made
transparent~\cite{BP,Pe}: the leading twist analysis describes the
$\Phi$ dissociation by gluons into a $Q$ and a $\bar Q$
\[
\Phi+g\to Q+\bar Q.
\]
To be energetically possible the gluon energy has therefore to be
larger than the $Q\bar Q$ Coulomb binding energy $\epsilon_0$. In view
of the fact that the confinement scale is small as compared to
$\epsilon_0$ the LT analysis then provides a description of $\Phi$
dissociation into open channels, e.g., $Q\bar q+\bar Q q$. Let us
emphasize that this dissociation is precisely the process of interest
for the question of $\Phi$ suppression in heavy ion collisions.

An important aspect for the phenomenology of the above cross section
is its limiting behaviors for both small and large energy
regimes. These are linked to the $x\to 1$ and $x\to 0$ behaviors of
$G$, respectively. It is then convenient to have in mind the simple,
yet standard, parameterization
\begin{equation}\label{eq:gluon_density}
G(x)=A\,(1-x)^\eta/x^{1+\delta}.
\end{equation}
With this ansatz one can write down exact asymptotic formulas either
by following the reasoning of Ref.~\cite{BP} or by noticing that the
$\Phi$-$h$ cross section is proportional to a hypergeometric
function. This is most easily done by changing variable $x$ to
$t=(x\lambda/\epsilon_0-1)/(\lambda/\epsilon_0-1)$ in
Eq.~(\ref{eq:im_amplitude}). Then one recognizes~\cite{AS}
\begin{eqnarray*}
\sigma(\lambda)&=&\frac{16^3\pi}{6N_c^2}
a_0^3\epsilon_0\,A\,B(\eta+1,5/2)(\lambda/\epsilon_0-1)^{\eta+5/2}\\
&&\times
(\lambda/\epsilon_0)^{\delta-\eta}
{}_2F_1(\delta+6,5/2;\eta+7/2;1-\lambda/\epsilon_0).
\end{eqnarray*}
For $\lambda$ in the neighborhood of $\epsilon_0$, i.e., the $\mh=0$
threshold, the hypergeometric function approaches 1 and we get
\begin{equation}\label{eq:threshold}
\sigma_{\Phi\,h}(\lambda)\sim
\frac{16^3\,\pi}{6N_c^2}a_0^3\epsilon_0\,A\,B(\eta+1,5/2)
\,(\lambda/\epsilon_0-1)^{\eta+5/2}.
\end{equation}
For large energies, using
\[
{}_2F_1(\delta+6,\frac{5}{2};\eta+\frac{7}{2};1-\lambda/\epsilon_0)
\sim\frac{B(\delta+7/2,5/2)}{B(\eta+1,5/2)}
(\lambda/\epsilon_0)^{-5/2},
\]
one obtains
\begin{equation}\label{eq:asymptotic}
\sigma_{\Phi\,h}(\lambda)\sim
\frac{16^3\,\pi}{6N_c^2}a_0^3\epsilon_0\,A\,B(\delta+7/2,5/2)
\left(\frac{\lambda}{\epsilon_0}\right)^{\delta}.
\end{equation}
The high energy cross section is primarily geometrical (remember
$a_0^3\epsilon_0\propto\alpha_S a_0^2$). In addition to this simple
behavior, there is a non trivial energy dependence coming from the
small $x$ behavior of the gluon density.

For phenomenological investigations we shall also use slightly more
involved forms for $G(x)$ as obtained in parton distribution function
studies. In this case the connection to ${}_2F_1$ is lost. One may,
however, derive similar asymptotic expressions by first expanding the
gluon distribution either in the neighborhood of 1 or 0.

\subsection{Massive target}\label{sub:mhneq0}

Having illustrated the method for the case $m_h=0$, we now turn to the
general case $m_h\neq 0$. Plugging Eq.~(\ref{eq:gluon_matrix_element})
into Eq.~(\ref{eq:lt_amplitude}) leads to
\begin{eqnarray*}
{\cal M}'(\lambda)&=&a_0^3\epsilon_0^2\sum_{k\ge 1} M_{2k}\\
&&\times\sum_{j=0}^k \frac{(2k-j)!}{j!(2k-2j)!}
(\lambda/\epsilon_0)^{2k-2j}
\left(-\frac{\mh^2}{4\epsilon_0^2}\right)^j,
\end{eqnarray*}
with $M_{2k}=d_{2k}A_{2k}$. We thus get an amplitude which may be
considered as a double power series in $\lambda$ and $\mh$. The study
of this double series with complex arguments $\lambda\to\cl$ and
$\mh^2/(4\epsilon_0^2)\to z$ shows that it is absolutely convergent
for $|\cl/\epsilon_0|+|z|<1$. In this domain, defining $k'=k-j$ we
may rewrite the series as
\begin{eqnarray}
{\cal M}'(\lambda,\mh)&=&a_0^3\epsilon_0^2
\sum_{j\ge 0,k'\ge 1}(\lambda/\epsilon_0)^{2k'}M_{2(k'+j)}
\frac{(2k'+j)!}{j!(2k')!}\nonumber\\
\label{eq:power_series_p}
&&\!\!\!\!\times\left(-\frac{\mh^2}{4\epsilon_0^2}\right)^j
+a_0^3\epsilon_0^2\sum_{j\ge 1}M_{2j}
\left(-\frac{\mh^2}{4\epsilon_0^2}\right)^j.
\end{eqnarray}

The second term on the right-hand side corresponds to the power-series
expansion of the scattering amplitude $\mathcal{M}$ of
Section~\ref{sub:mh0} evaluated at the complex plane location
$\cl=i\mh/2$:
\[
a_0^3\epsilon_0^2\sum_{j\ge 1}M_{2j}
\left(-\frac{\mh^2}{4\epsilon_0^2}\right)^j
=\mathcal{M}(i\mh/2).
\]
{}From the representation~(\ref{eq:amplitude}) of the scattering
amplitude we immediately see that $\mathcal{M}(i\mh/2)$ is well
defined and real for every (real) $\mh$. We thus ignore this term in
the following since it does not contribute to the total cross section
at leading twist.

Let us now concentrate on the first term on the right-hand side of
Eq.~(\ref{eq:power_series_p}). We use the same reasoning as in
Section~\ref{sub:mh0} considering now the double
series~(\ref{eq:power_series_p}) with complex arguments $\cl$ and $z$
instead of $\lambda$ and $\mh^2/(4\epsilon_0^2)$. Expressing first
$M_n$ as the $n$-th moment of $h=G\otimes f$ one can write
\begin{eqnarray*}
M_{2(k'+j)}\frac{(2k'+j)!}{j!(2k')!}(-z)^j&=&\int_0^1\frac{dx}{x} x^{2k'}
h(x)\\
&&\times\frac{(2k'+j)!}{j!(2k')!} (-x^2\,z)^j.
\end{eqnarray*}
In the convergence domain of the double series, the series in $j$ may
be summed up. Introducing
\begin{eqnarray}\label{eq:moments_p}
M'_n(z)&=&\sum_{j\ge 0}M_{n+2j} \frac{(n+j)!}{j!n!} (-z)^j\\
\nonumber
&=&\int_0^1 \frac{dx}{x}x^n h(x)\frac{1}{(1+x^2\,z)^{n+1}},
\end{eqnarray}
we may then follow another time the reasoning of Section~\ref{sub:mh0}
replacing $M_n$, Eq.~(\ref{eq:moments}), by $M'_n(z)$ as given by
Eq.~(\ref{eq:moments_p}). We notice in passing that the modified
moments $M'_n(z)$ are analytic functions of $z$ throughout the complex
plane except at $z=-1$.

For simplicity we restrict ourselves to physical masses, i.e., to the
positive real axis of $z$ where the above integral representation is
well defined. Then, two cases show up depending on whether
$\mh<2\epsilon_0$ or not. The former case is the one relevant to
phenomenology but we consider both cases in turn for completeness.

For $\mh<2\epsilon_0$, we restore $z=\mh^2/(4\epsilon_0^2)$ and change
the variable to $x'=x/(1+x^2 \mh^2/(4\epsilon_0^2))$. Then we have
\[
M'_n\left(\frac{\mh^2}{4\epsilon_0^2}\right)
=\int_0^{[1+\mh^2/(4\epsilon_0^2)]^{-1}}
\frac{dx'}{x'}x'^n\frac{h(x)}{1-x^2\,\mh^2/(4\epsilon_0^2)},
\]
with $x$ understood as a function of $x'$. In this form we can easily
follow the reasoning of Section~\ref{sub:mh0} because the integration
range does not play a role until one cuts the amplitude. This cutting
imposes $x'=\epsilon_0/\lambda$ and thus results in a non vanishing
imaginary part for
$\lambda>\lambda_0=\epsilon_0+\mh^2/(4\epsilon_0)$. Above this
threshold
\begin{eqnarray}
\im\mathcal{M}'(\lambda)&=&\frac{\pi}{2}a_0^3\epsilon_0^2\,
\frac{h(\epsilon_0/\lambda_+)}{1-\mh^2/(4\lambda_+^2)}
\nonumber\\
\label{eq:im_amplitude_p}
&=&\frac{\pi}{2}a_0^3\epsilon_0^2\,
\frac{\lambda_+}{\sqrt{\lambda^2-\mh^2}}\,
h(\epsilon_0/\lambda_+),
\end{eqnarray}
where $\lambda_+=\Big(\lambda+\sqrt{\lambda^2-\mh^2}\Big)/2$. Dividing
Eq.~(\ref{eq:im_amplitude_p}) by the flux factor we obtain for the
total cross section
\begin{equation}\label{eq:cross_section_p}
\sigma_{\Phi\,h}(\lambda)=\frac{\lambda_+^2}{\lambda^2-\mh^2}
\int_0^1 dx\,G(x)\,\sigma_{\Phi\,g}(x\lambda_+).
\end{equation}

Some comments are in order. We first stress that, as in the $\mh=0$
case, one arrives at a simple partonic form of the cross section. This
means, in particular, that the physical discussion we gave after
Eq.~(\ref{eq:cross_section}), based on the subprocess $\Phi+g\to
Q+\bar Q$, still holds in the massive target case. Apart from the
prefactor, the only modification between
Eq.~(\ref{eq:cross_section_p}) and Eq.~(\ref{eq:cross_section}) is the
modification of the expression for the gluon energy in the partonic
cross section $\sigma_{\Phi\,g}$. The change from $x\lambda$ to
$x\lambda_+$ may be given a heuristic interpretation in the parton
context using light cone coordinates\footnote{Such a connection
between light cone variables and $\mh\neq 0$ correction is discussed
for deep inelastic scattering in~\cite{DGP}.}. In the $\Phi$ rest
frame let us choose the third axis along the hadron $h$ momentum and
form $p^+=(\lambda+\sqrt{\lambda^2-\mh^2})/\sqrt{2}$ and
$p^-=\mh^2/(2p^+)$. In the parton picture the gluon causing the
dissociation is picked up from the hadron $h$ and has a negligible
transverse momentum, and hence a negligible minus momentum. Its energy
is then easily expressed in terms of $x$, the light cone (plus)
momentum fraction of the gluon, and reads $\omega=xp^+/\sqrt{2}$,
i.e., $\omega=x\lambda_+$.

Our understanding of the prefactor is more formal. We observe that
Eq.~(\ref{eq:im_amplitude_p}) may be rewritten as
\[
\im\mathcal{M}'(\lambda)\,d\lambda
=\im\mathcal{M}(\lambda_+)\,d\lambda_+,
\]
a relation which entails the (formal) identity
\begin{equation}\label{eq:-1_sumrule}
M'_{-1}(z)=M_{-1},\quad\forall z,
\end{equation}
which can also be obtained from a direct comparison between
Eqs.~(\ref{eq:moments_p}) and~(\ref{eq:moments}).

Next, we point out that, as expected, the $\mh\neq 0$ corrections are
sizeable only for small energies. The first aspect of these
corrections is that, as above mentioned, the threshold is now located
at
\begin{equation}\label{eq:threshold_shift}
\lambda_0=\epsilon_0+\frac{\mh^2}{4\epsilon_0}.
\end{equation}
As in the massless case this corresponds to the need to find in $h$ a
gluon with an energy $\omega\ge\epsilon_0$ sufficient to dissociate
the $\Phi$. With $\omega=x\lambda_+$ and $x\le 1$ this gives a
$\lambda_+$ threshold $\lambda_{+0}=\epsilon_0$, leading to
Eq.~(\ref{eq:threshold_shift}). The second aspect is that the cross
section behavior for $\lambda\to\lambda_0$ is given by
Eq.~(\ref{eq:threshold}) with an argument $\lambda_+$ instead of
$\lambda$ and a prefactor
\[
\frac{\lambda_+^2}{\lambda^2-\mh^2}
\sim\left(\frac{\epsilon_0}{\epsilon_0-\mh^2/(4\epsilon_0)}\right)^2.
\]

Let us now investigate the $\mh>2\epsilon_0$ case. One may perform the
same change of variable in the intervals $[0,2\epsilon_0/\mh]$ and
$[2\epsilon_0/\mh,1]$ leading to
\begin{eqnarray*}
M'_{n'}(\mh^2/(4\epsilon_0^2))=
\int_0^{\epsilon_0/\mh}\frac{dx'}{x'}x'^{n'}\frac{h(x)}
{1-x^2\,\mh^2/(4\epsilon_0^2)}&&\\
+\int_{[1+\mh^2/(4\epsilon_0^2)]^{-1}}^{\epsilon_0/\mh}
\frac{dx'}{x'}x'^{n'}\frac{h(x)}{x^2\,\mh^2/(4\epsilon_0^2)-1}.&&
\end{eqnarray*}
The threshold becomes $\lambda_0=\mh$ and one finds 
\begin{equation}\label{eq:im_amplitude_s}
\im\mathcal{M}'(\lambda)=\frac{\pi}{2}a_0^3\epsilon_0^2
\left(\frac{h(\epsilon_0/\lambda_+)}{1-\mh^2/4\lambda_+^2}+
\frac{h(\epsilon_0/\lambda_-)}{\mh^2/4\lambda_-^2-1}\right),
\end{equation}
with $\lambda_\pm=\Big(\lambda\pm\sqrt{\lambda^2-\mh^2}\Big)/2$. We
notice that the first contribution is the one already obtained in the
case $\mh<2\epsilon_0$. The second term is new but contributes only in
the range $\mh\le\lambda\le\epsilon_0+\mh^2/(4\epsilon_0)$. We further
point out that the relevant energy variable is now half the difference
between energy and momenta, instead of half the sum for the first
term, and that the present result Eq.~(\ref{eq:im_amplitude_s}) is
again consistent with the $M_{-1}$ sum
rule~(\ref{eq:-1_sumrule}). Finally, in the neighborhood of threshold
the cross section now amounts to
\begin{equation}\label{eq:threshold_s}
\sigma_{\Phi\,h}(\lambda)\sim\frac{1}{2(\lambda^2/\mh^2-1)}
\frac{\pi}{2}a_0^3\epsilon_0
\frac{h(2\epsilon_0/\mh)}{\mh/2\epsilon_0}.
\end{equation}

\section{Phenomenology}

\subsection{Choice of parameters}\label{sub:choice}

In view to give numerical values for $\Phi$-$h$ leading twist total
cross sections and thresholds, it is necessary to fix, on the one
hand, the heavy quark mass $m_Q$ and the quarkonium Rydberg energy
$\epsilon_0$, and, on the other hand, the gluon density in the target.

\subsubsection{Quarkonium sector}

The above described QCD analysis assumes that the $Q\bar{Q}$ binding
potential is well approximated by the Coulomb part of the QCD
potential~\cite{Pe}. Treating the 1S and 2S heavy quarkonia as
Coulombic states leads to
\begin{eqnarray}\label{1S-mass}
M_{Q\bar{Q}}(1S)&=&2 m_Q - \epsilon_0,\\
\label{2S-mass}
M_{Q\bar{Q}}(2S)&=&2 m_Q - \frac{\epsilon_0}{4},
\end{eqnarray}
that is
\begin{eqnarray*}
m_Q&=&\frac{1}{6}\left[4\,M_{Q\bar{Q}}(2S)-M_{Q\bar{Q}}(1S)\right],\\
\epsilon_0&=&\frac{4}{3}\left[M_{Q\bar{Q}}(2S)-M_{Q\bar{Q}}(1S)\right].
\end{eqnarray*}
This gives for charm and bottom respectively (set (i))
\begin{eqnarray*}
\epsilon_{0\,c}&=&0.78\;\mathrm{GeV},\quad m_c = 1.94\;\mathrm{GeV},\\
\epsilon_{0\,b}&=&0.75\;\mathrm{GeV},\quad m_b = 5.10\;\mathrm{GeV}.
\end{eqnarray*}

One way of estimating the applicability of the heavy quark analysis to
charmonia and bottomonia is to compare the size of each Coulomb-state
to typical confining distances. One may first evaluate the Bohr radius
$a_0=1/\sqrt{\epsilon_0 m_Q}$, this gives $a_{0\,c}=0.16$~fm for charm
and $a_{0\,b}=0.10$~fm for bottom. Recalling that the 1S-state root
mean square is given by $r(1S)=\sqrt{3}\,a_0$ one finds that the
1S-state size remains somewhat below typical confining distances. We
then consider that the LT analysis may be at least indicative of the
behavior of 1S-state cross section. Computing the size of 2S-states
with $r(2S)=\sqrt{30}\,a_0$, one sees that the situation is much less
favorable for 2S-states, especially for charmonium. The application
of the framework to $\Upsilon'$ is given in
Appendix~\ref{app:2S_states}.

In addition to the question of the validity of the computation of
2S-state cross section within the LT analysis this also led us to
reconsider the above choice of parameters. For this we drop
Eq.~(\ref{2S-mass}) and propose to fix the Rydberg energy to the
energy-gap between the 1S-state and the open flavor production. With
Eq.~(\ref{1S-mass}) this is equivalent to putting $m_c=m_D$ for charm
and $m_b=m_B$ for bottom. We then have the alternative set (set~(ii))
\begin{eqnarray*}
\epsilon_{0\,c}&=&0.62\;\mathrm{GeV},\quad m_c = 1.86\;\mathrm{GeV},\\
\epsilon_{0\,b}&=&1.10\;\mathrm{GeV},\quad m_b = 5.28\;\mathrm{GeV}.
\end{eqnarray*}
As we shall see, the cross section is only sizeable at large
energy. In this region the magnitude of the cross section is driven by
the factor $a_0^3\epsilon_0=1/\sqrt{m_Q^3 \epsilon_0}$ (see
Eq~(\ref{eq:asymptotic})). Taking set (ii) instead of set (i) results
in a 20\%~cross section increase for charm and in a 22\%~cross section
decrease for bottom. It turns out that this uncertainty is smaller
than the one coming from scale fixing discussed in the next
section. We will therefore limit our further considerations to cross
sections obtained using set (i).

\subsubsection{Gluon distributions}

The other important input for the computation is the gluon density
$G_h$ in the hadron $h$ considered. At this point one should remember
that this density depends on the factorization scale $\mu$ (see
paragraph following Eq.~(\ref{eq:lt_amplitude})). In the $\Phi$-$h$
cross section, not only $G_h$ is a function of $\mu$, but also
$\sigma_{\Phi\,g}$. Part of the dependence stands in the explicit
coupling of the gluon with the $\Phi$ constituents, corresponding to
the factor $\alpha_S$ in $a_0^3\epsilon_0\propto\alpha_S a_0^2$,
though the knowledge of the full dependence requires a complete
one-loop calculation. Lacking such an analysis led us first to restrict
ourselves to the so-called leading order (LO) gluon distributions
(i.e., their evolutions are computed to one-loop) and correspondingly
to the LO~running coupling. Second, we investigated with some care the
variation of our results with the choice of different factorization
scales.

We started this analysis with the prescription suggested in
Ref.~\cite{Pe} that $\mu\sim\epsilon_0$. For such a low scale the only
available parameterization for the proton is that of
Gl\"uck-Reya-Vogt~\cite{GRV94}, labelled GRV94~LO\footnote{It should
be borne in mind that at such a low normalization scale the gluon
distribution is less constrained by experimental data than by model
assumptions.}. These authors have proposed a parameterization for the
pion too~\cite{GRV92}. Concerning the $x$ dependence of the various
gluon densities, one should notice that the intermediate-$x$ region is
rather well monitored, while the small-$x$ region is poorly
understood, especially for the pion case. Thanks to the HERA
measurements the situation for the proton is much better. Including
these data Gl\"uck-Reya-Vogt have provided a new
parameterization~\cite{GRV98} but its lower scale is larger than that
needed for this study. However we checked that at a scale large enough
for both to be compared the difference between the gluon distribution
of Ref.~\cite{GRV98} and that of Ref.~\cite{GRV94} is not significant.

We also examined, at a larger factorization scale, the consequences of
different parameterizations of the gluon distribution in the proton,
considering in turn MRST98~LO~\cite{MRST} and CTEQ5L~\cite{CTEQ5}.

\subsection{Cross section variation with energy including target mass
corrections}\label{sub:variation}

As we shall see in the next section, the cross section at a given
energy depends on the choice of parameters. Its general trend,
however, is rather independent of a specific choice. We therefore
begin our phenomenological study by discussing in this section those
aspects that only weakly depend on the quarkonium parameters and gluon
distributions.

In Sect.~\ref{sec:derivation} we have seen that one can distinguish
two extreme energy regimes in the $\Phi$-$h$ cross section: a
threshold region and a high energy regime. The high-energy cross
section is independent of the target mass and is given by
\[
\sigma(\lambda)\sim\sigma_{\mathrm{as}}(\lambda)
=C\,(\lambda/\epsilon_0)^\delta
\]
for gluon densities $G(x)\sim\mathrm{const.}/x^{1+\delta}$ at small
$x$. The constant $C$ depends on details of the gluon density and on
the parameters describing the charmonium sector (see, e.g.,
Eq.~(\ref{eq:asymptotic})). In every case discussed in
Sect.~\ref{sub:choice} the relative difference between this asymptotic
cross section $\sigma_{\mathrm{as}}$ and the full result is less than
25\% for $\lambda > 30\,\epsilon_0$. For $J/\psi$ this translates
roughly into $\sqrt{s}>13$~GeV and for $\Upsilon$ into
$\sqrt{s}>23$~GeV.

For $\mh=0$ the cross section is very small in the threshold
region. We found that it is less than 10\% of $\sigma_{\mathrm{as}}$
for $\lambda<2$--3$\,\epsilon_0$, i.e., $\sqrt{s}<5$~GeV for $J/\psi$
and 12~GeV for $\Upsilon$.

\begin{figure}[t]
\begin{center}
\includegraphics[width=8.3cm]{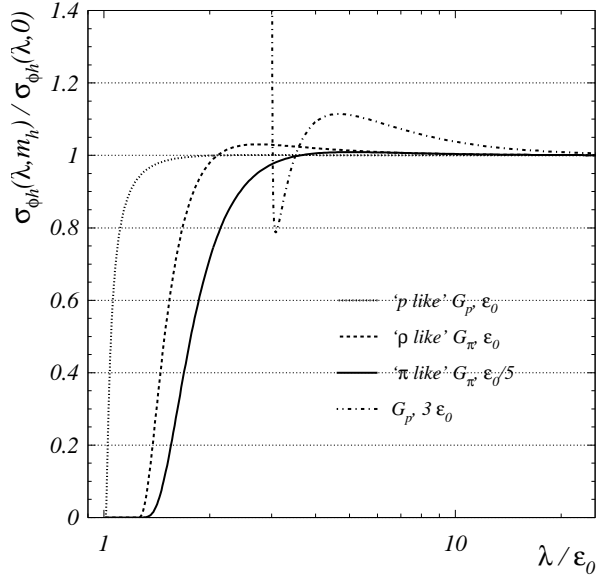}
\caption{Ratio of the corrected ($\sigma_{\Phi\,h}(\lambda,\mh)$) over
the uncorrected ($\sigma_{\Phi\,h}(\lambda,0)$) cross sections as a
function of $\lambda/\epsilon_0$. Calculations are performed for
hadron masses $\mh=\epsilon_0/5$, $\epsilon_0$ and $\mh=\epsilon_0$,
$3\,\epsilon_0$ using the gluon density $G_h(x,\mu=0.75~\mathrm{GeV})$
in the pion~\cite{GRV92} and in the
proton~\cite{GRV94}, respectively (see text).}
\label{fig:xs0xs}
\end{center}
\end{figure}

Let us now investigate the effect of a finite
$\mh$. Figure~\ref{fig:xs0xs} shows the ratio of
$\sigma_{\Phi\,h}(\lambda,\mh)$ and $\sigma_{\Phi\,h}(\lambda,0)$ for
two hypothetical hadron masses such that $\mh<2\epsilon_0$, namely
$m_h=\epsilon_0/5$ (``pion-like'', \textit{dotted}) and
$m_h=\epsilon_0$. The latter case has been computed for a pion gluon
distribution (``rho-like'', \textit{dashed}) as well as for a proton
gluon distribution (``proton-like'', \textit{solid}). This ratio is
plotted as a function of $\lambda/\epsilon_0$ and is then identical
for charm and bottom mesons.

In addition to the shift of the threshold
Eq.~(\ref{eq:threshold_shift}), we observe that the inclusion of
finite mass correction reduces the cross section close to
threshold. This result is opposite to what is found in
Ref.~\cite{KSSZ}\footnote{We give ample details on the comparison
between the present approach and that of Ref.~\cite{KSSZ} in
Appendix~\ref{app:sum_rules}.}, where it is argued that the target
mass correction tends to increase the $J/\psi$-$p$ cross section near
threshold. We notice that the mass correction is important only for
$\lambda <2$--$3\,\epsilon_0$. This implies that it is of limited
phenomenological interest since, as we have seen above, the cross
section is very small in this low energy region.

In Sect.~\ref{sub:mhneq0} we identified a different behavior in the
case of heavy targets ($\mh>2\,\epsilon_0$). Figure~\ref{fig:xs0xs}
shows mass correction for a hadron with mass $\mh =3\,\epsilon_0$ and
a gluon distribution $G_p$ given by Ref.~\cite{GRV94}
(\textit{dash-dotted}). The cross section diverges at threshold
($\lambda_0/\epsilon_0=\mh/\epsilon_0=3$), as can be seen in
Eq.~(\ref{eq:threshold_s}). The window for which the cross section
gets sizeable is very narrow, however. Thus this threshold behavior has
probably very little phenomenological implications.

\subsection{$\Phi$-$h$ absolute cross sections}

\subsubsection{Factorization scale dependence}

Before addressing in the next section the magnitude of the cross
sections, the influence of the factorization scale is quantitatively
investigated. More specifically the three choices:
$\mu^2=\epsilon_0^2$, $2\,\epsilon_0^2$, and $4\,\epsilon_0^2$ were
considered.

We made this investigation for the bottom channel with the parameter
set (i). In addition to the heavy quark mass, this choice benefits
from the numerical coincidence that\footnote{For consistency we use
here and in the following the one-loop running coupling with $n_f=3$
and with the QCD scale determined by Refs.~\cite{GRV94,GRV92}:
$\Lambda^{(3)}=232$~MeV.}
\[
a_{0\,b}\epsilon_{0\,b}=\sqrt{\frac{\epsilon_{0\,b}}{m_b}}
\approx\frac{2}{3}\,\alpha_S(\mu^2=\epsilon_{0\,b}^2),
\]
as expected for a large enough $\epsilon_0$ if the factorization scale
is $\mu^2=\epsilon_0^2$.

Let us first define the $\Upsilon$-$p$ cross section,
$\sigma^{(\mu^2=\epsilon_0^2)}_{\Upsilon\,p}$, computed with the
prescription $\mu^2=\epsilon_0^2$, to be that given by
Eq.~(\ref{eq:cross_section}) with the parameter set (i) for the
bottomonium and the proton distribution GRV94~LO evaluated at
$\mu^2=\epsilon_0^2$. We next define $\sigma^{(\mu^2)}_{\Upsilon\,p}$
at another scale to be that computed with GRV94~LO evaluated at
$\mu^2$ and multiplied by the factor
\[
\alpha_S(\mu^2)/\alpha_S(\epsilon_0^2),
\]
which takes into account the change of the coupling of the gluon to
the $\Phi$ constituents, and decreases down to 63\% for
$\mu^2=4\,\epsilon_0^2$.

\begin{figure}[t]
\begin{center}
\includegraphics[width=8.3cm]{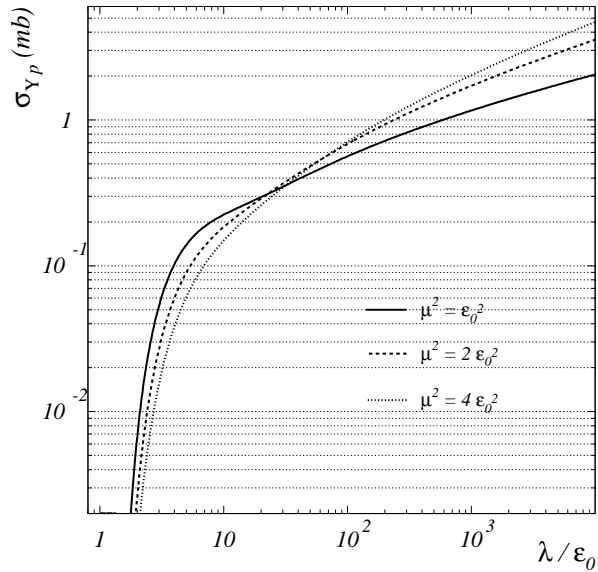}
\caption{Absolute $\sigma^{(\mu^2)}_{\Upsilon p}$ cross section as a
function of $\lambda/\epsilon_0$ for different factorization scale:
$\mu^2=\epsilon_0^2$ (\textit{solid}), $\mu^2=2\,\epsilon_0^2$
(\textit{dashed}), and $\mu^2=4\,\epsilon_0^2$ (\textit{dotted}). The
gluon distribution used is GRV94~LO.}
\label{fig:scale}
\end{center}
\end{figure}

On Figure~\ref{fig:scale} is shown the energy dependence of the
$\Upsilon$-$p$ cross sections evaluated with the GRV94~LO gluon
distribution at scales $\mu^2=\epsilon_0^2$ (\textit{solid}),
$2\,\epsilon_0^2$ (\textit{dashed}), and $4\,\epsilon_0^2$
(\textit{dotted}). To bypass the question of the scale dependence of
$\epsilon_0$ we restricted our study to the massless target case and
studied the cross section as a function of $\lambda/\epsilon_0$.

We first remark that the higher the scale $\mu^2$, the larger
(resp. smaller) the cross section at high (resp. low) incident
energy. At high energy ($\lambda/\epsilon_0 \sim 10^4$) the
uncertainty may be as high as a 100\%. The situation is much better in
the range $\lambda/\epsilon_0 \sim 20-100$ that is particularly
relevant for phenomenology. We also notice that the running of
$\alpha_S$ and that of $G_p$ tends to somewhat compensate each other
for the cross section at high energy.

\begin{figure}[t]
\begin{center}
\includegraphics[width=8.3cm]{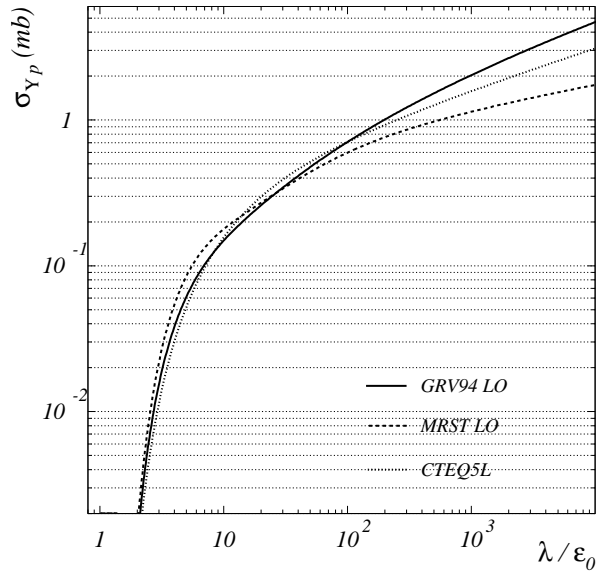}
\caption{Absolute $\sigma_{\Upsilon p}$ cross section as a function of
$\lambda/\epsilon_0$ using GRV94~LO (\textit{solid}), MRST98~LO
(\textit{dashed}), and CTEQ5L (\textit{dotted}) gluon
distribution. The scale has been fixed for all distributions to
$\mu^2=4\,\epsilon_0^2$.}
\label{fig:pdf}
\end{center}
\end{figure}

We also investigated the consequences of changing the parameterization
of the gluon distributions. Three leading order sets exist which may
be evaluated at the scale
$\mu^2=4\,\epsilon_0^2$. Figure~\ref{fig:pdf} displays the energy
dependence of the absolute cross section $\sigma_{\Upsilon\,p}$ using
GRV94~LO (\textit{solid}), MRST98~LO (\textit{dashed}) and CTEQ5L
(\textit{dotted}) gluon densities. The energy dependence proves to be
rather independent of a specific choice for energies
$\lambda<300\,\epsilon_0$. At larger energies there is a rather strong
dependence which leads to an uncertainty on the cross section
comparable to that due to the scale variation (compare
Figs.~\ref{fig:scale} and~\ref{fig:pdf}). The origin of this
uncertainty is the poor knowledge of the gluon distribution at very
low $x$.

\subsubsection{Cross sections using GRV gluon distributions}

\begin{figure}[t]
\begin{center}
\includegraphics[height=8cm]{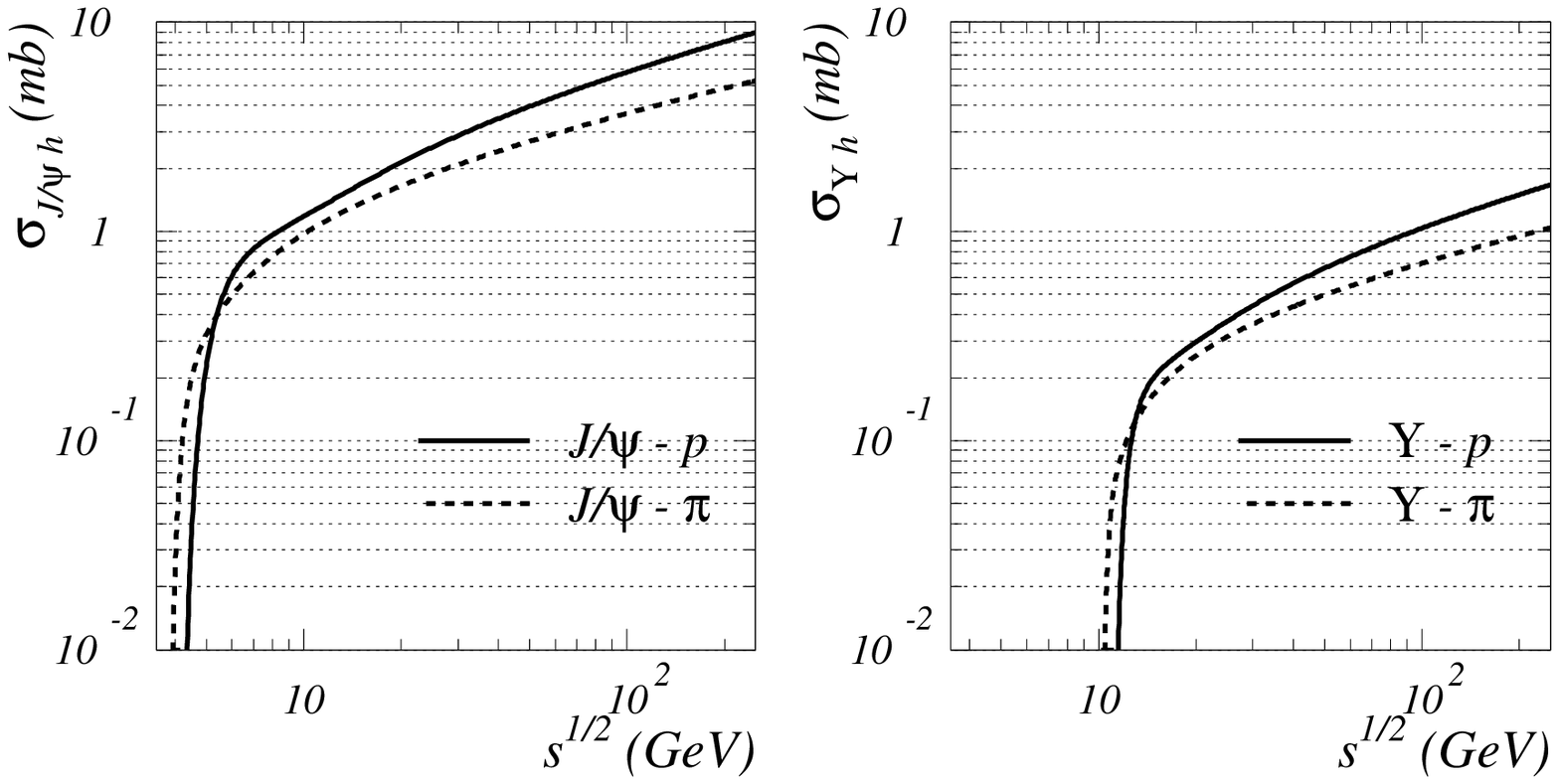}
\caption{Absolute cross sections $\sigma_{\Phi\,h}$ as a function of
the incident energy for $J/\psi$ (\textit{left}) and $\Upsilon$
(\textit{right}) with proton (\textit{solid}), and $\pi$
(\textit{dashed}). The gluon distributions $G_p(x)$ and $G_{\pi}(x)$
used come from Ref.~\cite{GRV94} and Ref.~\cite{GRV92}
respectively, evaluated at $\mu^2=\epsilon_0^2$.}
\label{fig:xsabs}
\end{center}
\end{figure}

We now turn to the discussion of the magnitude of the cross
section. The $J/\psi$ and $\Upsilon$ cross sections are displayed on
Fig.~\ref{fig:xsabs}. They have been computed using the parameter
set~(i) with the GRV gluon density for the proton~\cite{GRV94} and the
pion~\cite{GRV92} evaluated at $\mu^2=\epsilon_0^2$.

These cross sections $\sigma_{J/\psi\,h}$ (\textit{left}) and
$\sigma_{\Upsilon\,h}$ (\textit{right}) are found to strongly increase
up to about 1~mb and 0.2~mb, respectively. The transition between low
and high energy is situated around $\sqrt{s_{J/\psi\,h}}=8$~GeV and
$\sqrt{s_{\Upsilon\,h}}=15$~GeV, respectively. We also notice that,
depending on the set of parameters chosen (resp. (i) and (ii)), the
ratio $\sigma_{J/\psi\,h}/\sigma_{\Upsilon\,h}$ at high energy lies in
the range 4--6, i.e., roughly the charm to bottom ratio of
$a_0^3\epsilon_0$.

The energy dependence of $\sigma_{\Phi\,\pi}$ turns out to be
remarkably similar to the one in the proton channel, with a slightly
smaller magnitude. This similarity is intimately related to the
analogy that exists between the proton and the pion distributions in
the GRV approach. Needless to say that lacking small $x$ experiments
for the pion it has not been possible to verify this analogy so
far. With GRV distributions and at high energy
($\sqrt{s_{\Phi\,h}}=200$~GeV) the ratio
$\sigma_{\Phi\,\pi}/\sigma_{\Phi\,p}\approx 0.6$, independent of the
quarkonium considered.

\section{Conclusion}

The operator product expansion analysis has been widely used in the
analysis of deep inelastic scattering. Subsequently, these very
techniques proved useful to investigate heavy quark
systems~\cite{Pe,BP} allowing the calculation of the $J/\psi$-$p$
cross section within perturbative QCD. Such a cross section is of
seminal importance in the context of heavy ion collisions.

The present study is a continuation of the work of Bhanot and Peskin
and of a more recent paper by Kharzeev and collaborators~\cite{KSSZ}.
Let us gather what have been carried out here.

First, the leading twist forward scattering amplitude has been given a
simple integral expression, entailing a partonic representation for
the total cross section. Such a description had been found in a
different way in Ref.~\cite{BP} in the case of massless targets.

Secondly, finite target mass corrections have been systematically
incorporated. We showed that the cross section still assumes a
partonic form though in terms of a modified energy variable. In
addition to a shift of the reaction threshold, finite mass corrections
add to the suppression of the cross section at low relative energy but
become insignificant far above threshold. In the case of heavy
targets, however, we noticed that the cross section becomes large just
above (and even diverges at) threshold.

Last, the energy dependence of $\sigma_{J/\psi\,h}$ and
$\sigma_{\Upsilon\,h}$ has been investigated for several targets. We
found that $\sigma_{\Phi\,\pi}$ and $\sigma_{\Phi\,p}$ are strongly
suppressed in the vicinity of the threshold. At large energy, the
cross section is proportional to $s^{\delta}$ for a target with
$G(x)\sim\mathrm{const.}/x^{1+\delta}$ at small $x$. With GRV gluon
distributions this leads to slowly rising cross sections for both
$\Phi$-$\pi$ and $\Phi$-$p$. However, we should emphasize that the
small-$x$ gluon distribution are not much constrained, especially that
of the pion. Indeed the weak control we have on the gluon
distribution, because of both the just mentioned poor small-$x$
knowledge and the sizeable scale dependence, turned out to be the main
source of uncertainty in the present approach. The $\Phi'$ cross
section have also been investigated, although the relevance of a
perturbative approach is not fully satisfied for 2S~states.

In addition to the hadron mass corrections considered here and beyond
perturbative corrections, higher twist corrections may also lead to
substantial modification for a not so heavy quark such as the
charm. Indeed we noticed in Sect.~\ref{sub:mh0} that the threshold
location may vary when one incorporates higher twist correction,
making finite a cross section which is zero at leading
twist. Considerations of this type of corrections is clearly outside
the scope of the present study. Part of these corrections may be
associated to the $\Phi$ sector and in particular to the confining
part of the heavy quark potential. Other corrections involving higher
twist operators in the hadron target have presumably their
counterparts in deep inelastic scattering.

As compared to the other approaches mentioned in the introduction the
$J/\psi$-$\pi$ cross section is very tiny at $\sqrt{s}=4$~GeV. In the
4--7~GeV energy range it strongly increases driven by the intermediate
$x$ region in the gluon distribution. This $x$ region is fairly under
control thanks to the momentum sum rule and consequently the
prediction, within the present approach, of cross section smaller than
1~mb below 7~GeV is rather robust.

A consequence of such a small cross section at small relative energies
is that destruction of $J/\psi$'s by comovers become very unlikely.
We are presently studying whether our results allows already for an
answer to the question whether a quark gluon plasma is formed
in ultra-relativistic heavy ion reactions. Our present approach is,
however, limited to the $J/\psi$ cross section, since for the excited
states the binding energy is not large as compared to the confining
scale.

\appendix

\section{Sum rules for $\sigma_{\Phi\,h}$}
\label{app:sum_rules}

In this first appendix, we establish the sum rules satisfied by the
leading twist total cross section $\sigma_{\Phi\,h}$. This allows us
to make contact with a similar derivation done in the case
$\mh=0$~\cite{BP} and $\mh\neq 0$~\cite{KSSZ}. The starting point is
the expression for the leading twist
amplitude~(\ref{eq:power_series_p}). Expressing $M_n=d_nA_n$ and
recalling that $A_n$ is the $n$-th moment of the gluon density in the
hadron target (see Sect.~\ref{sub:framework}), we may write the series
as
\begin{eqnarray}
{\cal M}'(\lambda,\mh)&=&a_0^3\epsilon_0^2\sum_{k'\ge 1}
(\lambda/\epsilon_0)^{2k'}\int_0^1\frac{dx}{x}x^{2k'}G(x)
\nonumber\\
\label{eq:series_int}
&&\times\sum_{j\ge0}\frac{(2k'+j)!}{j!(2k')!}d_{2k'+2j}
\left(-\frac{\mh^2}{4\epsilon_0^2}\,x^2\right)^j
\\
&&+a_0^3\epsilon_0^2\int_0^1\frac{dx}{x}G(x)\sum_{j\ge 1}d_{2j}
\left(-\frac{\mh^2}{4\epsilon_0^2}\,x^2\right)^j.\nonumber
\end{eqnarray}
Using the identities~\cite{AS}
\begin{eqnarray*}
&&\sum_{j\ge0} \frac{(n+j)!}{j!n!}
B\left(n+2j+\frac{5}{2},\frac{5}{2}\right)z^j=
B\left(n+\frac{5}{2},\frac{5}{2}\right)\\
&&\times\,{}_3F_2\left(\frac{5}{4}+\frac{n}{2},\frac{7}{4}
+\frac{n}{2},1+n;\frac{5}{2}+\frac{n}{2},3+\frac{n}{2};z\right),\\
&&\sum_{j\ge0}B\left(2j+\frac{9}{2},\frac{5}{2}\right)z^j=
B\left(\frac{9}{2},\frac{5}{2}\right)\\
&&\times\,{}_3F_2
\left(1,\frac{9}{4},\frac{11}{4};\frac{7}{2},4;z\right),
\end{eqnarray*}
the sums over $j$ in Eq.~(\ref{eq:series_int}) leads to
\begin{eqnarray}
{\cal M}'(\lambda,\mh)&=&a_0^3\epsilon_0^2
\sum_{k'\ge1}d_{2k'}\tilde{A}(2k')
\left(\lambda/\epsilon_0\right)^{2k'}\nonumber\\
\label{eq:series_mh}
&&- a_0^3\epsilon_0^2 \left(\frac{\mh^2}{4\epsilon_0^2}\right)
d_2\int_0^1dx\,x G(x)\\
&&\times\,{}_3F_2\left(1,\frac{9}{4},\frac{11}{4};
\frac{7}{2},4;-\frac{\mh^2}{4\epsilon_0^2}\,x^2\right),\nonumber
\end{eqnarray}
where
\begin{eqnarray*}
&&\tilde{A}(n)=\int_0^1\frac{dx}{x}x^{n}G(x)\\
&&\times\,{}_3F_2\left(\frac{5}{4}
+\frac{n}{2},\frac{7}{4}+\frac{n}{2},1+n;\frac{5}{2}
+\frac{n}{2},3+\frac{n}{2};-\frac{\mh^2}{4\epsilon_0^2}\,x^2\right).
\end{eqnarray*}
We point out that Eq.~(\ref{eq:series_mh}) is equivalent to the
equation (15) of~\cite{KSSZ} with slightly different notations.

As noted in~\cite{BP}, performing the integral
\[
\oint\frac{d\lambda}{2i\pi}\lambda^{-2\ell-1}{\cal M}'(\lambda,\mh).
\]
around a counterclockwise contour enclosing the origin gives the
coefficient of $\lambda^{2\ell}$ in the
amplitude~(\ref{eq:series_mh}). Wrapping the contour around the $u$
and $s$ channel cuts, the contour at infinity giving no
contribution\footnote{Both the singularity pattern and asymptotic
behavior of the scattering amplitude, which are necessary for the
present construction to hold, are best studied within the approach of
Sect.~\ref{sec:derivation}. In Ref.~\cite{BP}, these properties where
assumed from general properties of the elastic scattering amplitude.},
\[
\frac{2}{\pi}\int_{\mh}^{+\infty}d\lambda\,\lambda^{-2\ell-1}
\im{\cal M}'(\lambda,\mh)=a_0^3\epsilon_0^2d_{2\ell}\tilde{A}(2\ell)
\,\epsilon_0^{-2\ell}.
\]
We put $\mh$ as a lower bound of the integral but this does not
presume of the exact location of the threshold $\lambda_0$
(necessarily greater than or equal to $\mh$) implicitly contained in
the set of sum rules (see below). Using the optical theorem one
finally gets the sum rules for the $\Phi$-$h$ cross section
\begin{equation}\label{eq:sumrules_l}
\int_{\mh}^{+\infty}\!\!d\lambda\,\lambda^{-2\ell-1}
\sqrt{\lambda^2-\mh^2}\sigma_{\Phi\,h}(\lambda)=
\frac{\pi}{2}a_0^3\epsilon_0^2\,d_{2\ell}\tilde{A}(2\ell)
\,\epsilon_0^{-2\ell},
\end{equation}
which is what Bhanot and Peskin found (Eq.~(3.10) in Ref.~\cite{BP})
in the limit of massless target. In order to compare these results
with~\cite{KSSZ}, we introduce the variable $y=\mh/\lambda$ to get
\begin{equation}\label{correct_sum_rules}
\int_0^1\!\!dy\,y^{2\ell-2}\sqrt{1-y^2}
\sigma_{\Phi\,h}(\mh/y)\!=\!\left(\frac{\mh}{\epsilon_0}\right)^{2\ell-1}
I(2\ell)\tilde{A}(2\ell),
\end{equation}
with
\[
I(n)=\frac{\pi}{2}a_0^3\epsilon_0\,d_{n},
\]
as introduced in \cite{KSSZ}. The relation (\ref{correct_sum_rules})
gives the set of sum rules that should replace that given by Eq.~(16)
of Ref.~\cite{KSSZ}, where the prefactor
$\left(\mh/\epsilon_0\right)^{2\ell-1}$ is missing.

In order to verify that the sum-rule formalism leads to the same
result as we obtained in the present study we need to solve
Eq.~(\ref{correct_sum_rules}) for $\sigma_{\Phi\,h}$. This may be done
using Laplace transform techniques. Eq.~(\ref{correct_sum_rules})
writes
\[
\int_{0}^{1}dy\, y^{n-2} \sqrt{1-y^2}\sigma_{\Phi\,h}(\mh/y)=
g(n),\quad n=2,4,\ldots,
\]
with
\[
g(n)=\left(\frac{\mh}{\epsilon_0}\right)^{n-1}I(n)\tilde{A}(n).
\]
Defining $x=-\ln y$ and $n'=n-1$, we get the relation
\begin{eqnarray*}
&&\int_{0}^{+\infty}dx\,\exp(-n' x)
\underbrace{\sqrt{1-\exp(-2x)}\sigma_{\Phi\,h}(\mh \exp(x))}_{=f(x)}
\\
&&=g(n'+1),\quad n'=1,3,\ldots,
\end{eqnarray*}
that is $g(n'+1)$ is the Laplace transform of $f(x)$.

The above relation can be uniquely extended to a complex argument
$\nu$ (at least for Re$\,\nu>\delta$, see below) instead of the listed
$n'$. We can then obtain $f(x)$ by inverting the Laplace transform
\begin{equation}
f(x)=\frac{1}{2 i \pi}\int_{\nu'_0-i\infty}^{\nu'_0+i\infty}
d\nu\exp(\nu x) g(\nu+1),
\end{equation} 
where $\nu'_0$ is an arbitrary real chosen so that the integration
contour is located ``at the right'' of all singularities of
$g(\nu+1)$. For gluon distribution behaving as $x^{-(1+\delta)}$ for
$x\to 0$, integrals of the type $\int_{0}^{1}dx\,x^{\nu} G(x)
{}_3F_{2}$ are well behaved provided Re$\,\nu>\delta$. As a consequence
we performed the inverse Laplace transform choosing $\nu'_0>\delta$.

Setting $\nu=\nu'_0+i u=\nu_0-1+i u$, and then $\lambda=\mh\exp(x)$,
we thus obtain
\begin{eqnarray*}
&&\sqrt{1-\exp(-2x)}\sigma_{\Phi\,h}(\mh \exp(x))\\
&&=\frac{1}{2\pi}
\int_{-\infty}^{+\infty}du\,\exp[(\nu_0-1+iu)x]g(\nu_0+iu),
\end{eqnarray*}
and then
\begin{eqnarray}
\sigma_{\Phi\,h}(\lambda)&=&
\frac{\lambda}{2\pi\sqrt{\lambda^2-\mh^2}}\int_{-\infty}^{+\infty}du
\nonumber\\
&&\quad\times\left(\frac{\lambda}{\mh}\right)^{\nu_0-1+iu}
g(\nu_0+iu)\nonumber\\
&=&\frac{\lambda}{2\pi\sqrt{\lambda^2-\mh^2}}
\int_{-\infty}^{+\infty}du\nonumber\\
&&\quad\times\left(\frac{\lambda}{\epsilon_0}\right)^{\nu_0-1+iu}
I(\nu_0+iu)\tilde{A}(\nu_0+iu),
\label{eq:inverse_sum_rule}
\end{eqnarray}  
for every $\nu_0>1+\delta$. Let us remark that the $\epsilon_0$ energy
scale appearing in the first factor of the integrand would be $\mh$
with the sum rules proposed in~\cite{KSSZ}. The difference between
these two results is important when one considers the $\mh\to 0$
limit, in which case the latter expression is ill-defined contrarily
to ours, given in Eq.~(\ref{eq:inverse_sum_rule}).

We carried out a numerical evaluation and found out that it reproduces
the results obtained in the main body of the paper. One critical point
in the comparison between the two approaches is the verification that
the threshold is located at the predicted value, that is, in
particular, that the numerical result is compatible with zero below
this threshold (a point which is far from evident when one looks at
Eqs.~(\ref{eq:inverse_sum_rule}) or~(\ref{eq:sumrules_l})).

\section{2S states}
\label{app:2S_states}

In this appendix, calculations of the $\Phi$ cross sections are
extended to the 2S~states $\Phi'$. The modification amounts to
replacing the 1S~coefficients $d_{n}^{(1S)}$ (denoted $d_{n}$ so far)
by the 2S~coefficients~\cite{Pe}
\[
d_n^{(2S)}=\frac{16^3}{3N_c^2}\frac{\Gamma(n+5/2)\Gamma(5/2)}
{\Gamma(n+7)}(16n^2+56n+75).
\]
Expressing $d_n^{(2S)}$ as a function of the 1S coefficients
\[
d_n^{(2S)}=4^n(4d_n^{(1S)}-24d_{n+1}^{(1S)}+36d_{n+2}^{(1S)}),
\]
allows one to get an integral representation for the 2S coefficients
\begin{equation}\label{eq:int_dn2s}
d_n^{(2S)}=4^n\int_0^1\frac{dx}{x}x^n f^{(2S)}(x),
\end{equation}
with
\[
f^{(2S)}(x)=\frac{16^3}{3N_c^2} x^{5/2}(1-x)^{3/2}(2-6x)^2,
\]
which is the ingredient needed to carry out the procedure outlined in
Sect.~\ref{sec:derivation}. The changes are that the function $h$ has
to be replaced by $h^{(2S)}=G\otimes f^{(2S)}$ and that it is now
evaluated at $\epsilon/\lambda,\epsilon/\lambda_+,$ or
$\epsilon/\lambda_-$ instead of
$\epsilon_0/\lambda,\epsilon_0/\lambda_+,$ or $\epsilon_0/\lambda_-$
where $\epsilon=\epsilon_0/4$ is the binding energy of the
2S~state. The partonic expression is thus similar with
\[
\sigma_{\Phi'\,g}(\omega)=16\frac{16^3\pi}{6N_c^2}a_0^3\epsilon_0
\frac{(\omega/\epsilon-1)^{3/2}(\omega/\epsilon-3)^2}
{(\omega/\epsilon)^7}\theta(\omega-\epsilon).
\]

The energy dependence $\sigma_{\Upsilon'\,p}(s)$ has been computed
using the parameter set (i). The cross section diverges at threshold
($\sqrt{s}\approx 11$~GeV) and decreases to a minimum of about 4~mb
at a center of mass energy 0.6~GeV above threshold. Then the cross
section increases smoothly ($\sigma_{\Upsilon'\,p}(s)\propto
s^\delta$) and reaches 30~mb by $\sqrt{s}\approx 200$~GeV. We notice
that at high energy the ratio
$\sigma_{\Phi'\,h}/\sigma_{\Phi\,h}\approx 20$ for all incident
hadrons. Since $r(2S)=\sqrt{10}\,r(1S)$, this result lies somewhat
above the geometrical expectation $r^2(2S)/r^2(1S)$.

We have already insisted on the fact that the LT perturbative analysis
is most likely not adequate to describe the $\psi'$ channel. In this
case, the cross section amounts to $\sigma_{\Psi'\,p}({\sqrt
s})=30$~mb at $\sqrt{s}=10$~GeV.

\end{document}